\documentclass[sigconf,natbib=true,anonymous=false]{acmart}
\usepackage{float}
\usepackage{algorithm}
\usepackage{algorithmic}
\usepackage{multirow}
\usepackage{subcaption}
\usepackage{enumitem}

\usepackage{cleveref}
\usepackage{balance}

\renewcommand\footnotetextcopyrightpermission[1]{}

\begin{document}

\title{Dual Contrastive Transformer for Hierarchical Preference Modeling in Sequential Recommendation}

\author{Chengkai Huang}
\authornote{Corresponding author.}
\email{chengkai.huang1@unsw.edu.au}
\affiliation{
  \institution{The University of New South Wales}
  \city{Sydney}
  \state{NSW}
  \country{Australia}
}

\author{Shoujin Wang}
\email{shoujin.wang@uts.edu.au}
\affiliation{
  \institution{University of Technology Sydney}
  \city{Sydney}
  \state{NSW}
  \country{Australia}
}

\author{Xianzhi Wang}
\email{XIANZHI.WANG@uts.edu.au}
\affiliation{
  \institution{University of Technology Sydney}
  \city{Sydney}
  \state{NSW}
  \country{Australia}
}

\author{Lina Yao}
\email{lina.yao@unsw.edu.au}
\affiliation{
  \institution{CSIRO’s Data 61 and UNSW}
  \city{Sydney}
  \state{NSW}
  \country{Australia}
}

\renewcommand{\shortauthors}{Chengkai Huang, Shoujin Wang, Xianzhi Wang, \& Lina Yao}

\begin{abstract}
 
Sequential recommender systems (SRSs) aim to predict the subsequent items which may interest users via comprehensively modeling users' complex preference embedded in the sequence of user-item interactions. However, most of existing SRSs often model users' single low-level preference based on item ID information while ignoring the high-level preference revealed by item attribute information, such as item category. Furthermore, they often utilize limited sequence context information to predict the next item while overlooking richer inter-item semantic relations. To this end, in this paper, we proposed a novel hierarchical preference
modeling framework to substantially model the complex low- and high-level preference dynamics for accurate sequential recommendation. Specifically, in the framework, a novel dual-transformer module and a novel dual
contrastive learning scheme have been designed to discriminatively learn users’ low- and high-level
preference and to effectively enhance both low- and
high-level preference learning respectively. In addition,  a novel semantics-enhanced context embedding
module has been devised to generate more informative context embedding for
further improving the recommendation performance. Extensive experiments on six real-world datasets have demonstrated both the superiority of our proposed method over the state-of-the-art ones and the rationality of our design.      

\end{abstract}


\maketitle

\section{Introduction}

Sequential Recommender Systems (SRSs) aim to predict the next item which may interest a user via modeling her/his dynamic and timely preference. Such preference is usually modeled through a sequence of historical user-item interactions. Due to their strength of well-capturing users' dynamic and timely preferences, SRSs are able to provide accurate and timely recommendations~\cite{IJCAISurvey}.      

In recent years, SRSs have attracted increasing attention from both academia and industry. Hence, a variety of SRS models including both shallow and deep models have been proposed to improve the performance of sequential recommendations. Specifically, Recurrent Neural Networks built on Gate Recurrent Units (GRU) have been employed to model the long- and short-term point-wise sequential dependencies over user-item interactions for next-item recommendations~\cite{GRU4Rec1,song2021next}. Convolutional Neural Network (CNN)~\cite{Caser}, self-attention~\cite{SARS,wang2018attention,wang2020hierarchical} and Graph Neural Network~\cite{wang2020modelling,zheng2022ddghm} models have been incorporated into SRSs for capturing more complex sequential dependencies (e.g., collective dependencies) for further improving the recommendation performance. However, despite the remarkable performance that has been achieved, some significant gaps still exist in existing SRS methods, which greatly limit the further improvement of the recommendation performance.

\begin{figure}[t]  
	\centering
	\includegraphics[width=\linewidth]{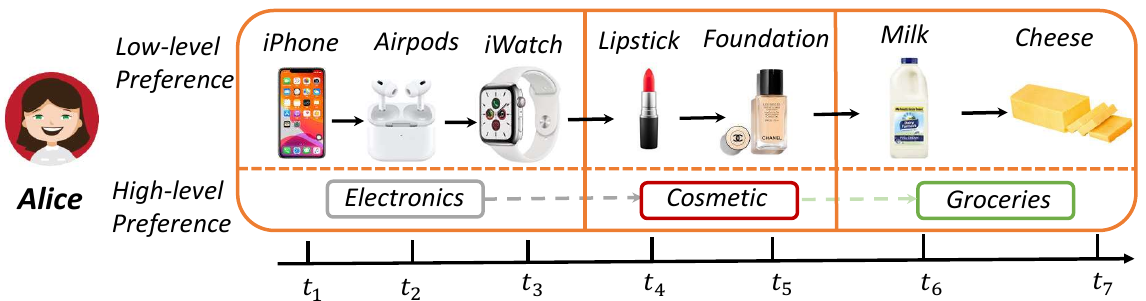}
	\caption{ Alice's hierarchical preference dynamics through a sequence of purchased items. Alice's low-level preference indicated by item ID changes sharply while her high-level preference indicated by category changes smoothly.
 }
	\label{fig:example}
\end{figure}

First, most of the existing SRS methods model a user's preferences by only relying on the low-level and specific ID information of items while overlooking the informative high-level signals, such as item category. \textit{However, \textbf{(Gap 1)} such a practice cannot accurately and comprehensively capture a user's complex hierarchical preference dynamics.} 
The reason is two-fold: (1) On one hand, a user's preference is essentially hierarchical with multi-granularity, including both \textit{high-level preference} towards different categories of items (e.g., Alice may like electronic products from "Apple" brand) and \textit{low-level preference} towards different items within each category (e.g., Alice may particularly like iPhone-13)~\cite{zhu2020sequential}. However, item ID information can indicate the low-level preference towards specific items only and thus the high-level preference has been ignored. (2) On the other hand, the changes of preference overtime at the low level are much faster than those at the high level. Such a difference is of great significance to precisely characterize a user's preference dynamics in SRSs but has been ignored by existing SRSs built on item ID information only. For instance, as shown in Figure \ref{fig:example}, 
when looking at Alice's successive purchases of iPhone and Airpods, her low-level preference has changed from one item to another, however, her high-level preference actually has not changed since she keeps focusing on the category of "Electronics".     

Second, the user-item interactions in the sequential recommendation are often limited and sparse, impeding the well learning of users' preferences. To alleviate this problem, Contrastive Learning (CL) has been introduced to SRSs to enhance user preference modeling by introducing more supervision signals via data augmentation \cite{S3-Rec, ContraRec}. \textit{However, \textbf{(Gap 2)} most of the CL-based SRSs only involve a single contrast based on the low-level preference indicated by item ID information, overlooking the contrast built on high-level preference indicated by item category information.} As a result, the high-level preference indicating users' relatively stable intention and demand may not be substantially learned, especially on highly sparse datasets ~\cite{CLproblem, CLproblem2}. In addition, the lack of contrast on the item categorical level may also lose some important constraint signal to connect (resp. distinguish) items from the same (resp. different) categories, further impeding the recommendation performance.   

Finally, in SRSs, the contextual information embedded in a user-item interaction sequence is the key signal to guide the prediction of the next item~\cite{GRU4Rec1,GRU4Rec2,wang2022sequential,liu2023collaborative,shuai2019}. \textit{However, in most existing SRSs, \textbf{(Gap 3)} such contextual information is often learned from item IDs without the consideration of richer semantic relations between items, resulting in un-informative context embedding and thus impeding next-item recommendations.} This triggers the urgent need for more effective context embedding to comprehensively capture complex item characteristics and inter-item relations in the sequence context.  

Aiming at bridging the aforementioned three significant gaps in existing SRS works, we propose a novel 
\textbf{Hierarchical Preference modeling} (HPM) framework for accurate sequential recommendations. In HPM, there are mainly three novel modules that are particularly designed to address the three gaps respectively. To be specific, to address the first gap, we design a \textbf{Dual-Transformer (DT) module} to comprehensively model both the low-level (i.e., item level) preference dynamics and high-level (i.e., category level) preference dynamics. 
To address the second gap, we propose a novel \textbf{Dual-Contrastive Learning (DCL) scheme} to better learn users' two-level preferences via the contrast on both the item level and category level.  

To bridge the third gap, we devise a novel \textbf{Semantics-enhanced Context Embedding Learning (SCEL) module} to well capture and incorporate the hidden semantic relations between items to generate more informative sequence context embedding for next-item prediction. Here, semantic relations refer to substitute/complementary relation between items, which are extracted from interaction data like co-clicked/co-purchased items by following common practice~\cite{chorus,SLRS,huang2023www}. These three modules are closely related and work collaboratively towards better users' hierarchical preference learning for accurate next-item prediction.    
 
The main contributions of this work are summarized below:
\begin{itemize}[leftmargin=*]
\item We propose modeling hierarchical preference dynamics for better-capturing users' timely and dynamic preferences for accurate sequential recommendations. Accordingly, we devise a novel hierarchical preference modeling (HPM) framework.         

\item We design a novel dual-transformer module and a novel dual contrastive learning scheme to equip the HPM framework. The former can discriminatively learn users' low- and high-level preferences while the latter can effectively enhance both low- and high-level preference learning without manually corrupting the original sequence data.

\item We propose a novel semantics-enhanced context embedding module to generate more informative context embedding for further improving the recommendation performance. 

\end{itemize}

\section{Related Work}

\subsection{Sequential Recommendation}

Sequential recommendation aims to leverage users' historical interactions to capture users’ dynamic preferences for next-item prediction. Rendle et al. \cite{FPMC} propose a first-order Markov chain-based sequential recommendation method via modeling the transitions between items over a sequence of baskets. 
After that, to capture high-order dependencies over items, He et. al.~\cite{FusingM} propose a higher-order Markov chain-based model for sequential recommendations. 
In recent years, benefiting the powerful capability of deep neural networks to capture the complex and dynamic dependencies embedded in sequences, a variety of deep learning-based sequential recommendation methods~\cite{Caser, GRU4Rec1, GRU4Rec2} have been proposed. For example, self-attention-based methods~\cite{SARS} have achieved very good performance via utilizing the transformer architecture to learn complex item-item relationships. Cen et. al. \cite{CenKDD20} have proposed a novel controllable multi-interest framework, which can capture multiple interests from user behavior sequences. Although great progress has been achieved in the area of sequential recommendation, most of the SRS methods only focus on learning users' low-level preferences towards different items based on item ID information. They generally ignore users' high-level preferences toward different types/categories of items. 
These works did not comprehensively model the multi-granular preferences of users and the different preference shift patterns at different levels. 

\subsection{Category-aware Preference Modeling}

Although deep learning-based SRS methods achieve impressive results, there are still only a few works focusing on modeling user hierarchical preference. 
Instead of using item IDs as only item attribute as prior solutions, recent methods begin to take the side information into consideration so as to better capture the user's preference.
Zhang et. al.~\cite{FDSN} propose the FDSA model, which combines two separate branches of self-attention blocks for item ID and side features and fuses them in the final stage.
Then, Liu et. al.~\cite{NOVA} propose the NOVA-SR model, which feeds both the pure item id representation and side information integrated representation to the attention layer, where the latter is only used to calculate the attention key and query and keeps the value non-invasive. 
Instead of using early fusion to get item representation, Xie et. al.~\cite{XieZK22} propose the DIF model, decoupling the attention calculation process of various side information to generate fused attention matrices for higher representation power.
Yuan et. al.~\cite{YuanDTSZ21} propose the ICAI-SR model, which utilizes the attribute-to-item aggregation layer before the attention layer to integrate side information into item representation with separate attribute sequential models.
Besides, Zhou et. al.~\cite{S3-Rec} propose to leverage self-supervised attribute prediction tasks in the pre-training stage.
However, recent works~\cite{CLproblem2} find that these implicit feature fusion methods do not well improve recommendation performance, especially on highly sparse datasets. 
In contrast, explicitly learning intent sequences and item sequences can improve sequential recommendation on sparse datasets.

\subsection{Contrastive Sequential Recommendation}

Due to its great success in the computer vision area~\cite{SimCLR}, contrastive learning (CL) has been widely introduced to more and more areas including recommender systems~\cite{ssl_survey}. The main idea of CL-based recommendation is to design an auxiliary task to enhance the recommendation performance by data augmentation. Specifically, Yao et. al.~\cite{SSL1} propose a self-supervised learning (SSL) framework for large-scale item recommendations, which uses both the masking and dropout methods to augment the original data. Zhou et. al.~\cite{S3-Rec} propose four different types of self-supervised tasks to enhance the recommendation model's generalizability. Chen et. al.~\cite{ICL} utilize the unlabelled user behavior sequences to learn the user's intent distribution functions and fuse it into the self-supervised learning framework for sequential recommendations. In addition, there are also some works focusing on the graph-based recommendation with node-level contrastive learning~\cite{DHCN, MHCN}. 
Although effective, these CL-based RSs often manually augment the data by changing the original data, which may introduce some unnecessary noise to mislead user preference learning and the subsequent recommendations. Meanwhile, these methods only use the item ID for contrastive learning, ignoring the constraints on the user's preferred item category, which may lead to a lack of category relevance in the recommended item list.

\section{Methodology}

\subsection{Problem Formulation}

Let $\mathcal{U}$ and $\mathcal{V}$ be the user set and item set respectively. Let the sequential recommendation dataset as $\mathcal{D}$, which contains each user-item interaction sequence in a certain timestamp. Then $\mathcal{D}=\{\mathcal{S}_1,...\mathcal{S}_i,...,\mathcal{S}_n\}$, where $S_i$ means the unique sequence for user $u_i \in U$. 
For each $u_i$, it is associated with a chronological sequence of items interacted by him/her, denoted as $\mathcal{S}_i=<\mathcal{O}_i, v_n>$, where $\mathcal{O}_i$ means the context information for $u_i$, $\mathcal{O}_i=\{ V_i, C_i, T_i \}$, 
$V_i=\{v_1,v_2,...,v_{n-1}\}$ denotes the item ID sequence, 
$C_i=\{c_1,c_2,...,c_{n-1}\}$ denotes the item category sequence and 
$T_i=\{t_1,t_2,...,t_{n-1}\}$ denotes each timestamp sequence. ($n$ is the length of the sequence.) 
The task of sequential recommendation is to predict the next item ID $v_t$ which may interest the user based on historical interaction information $\mathcal{O}_i$. Specifically, for each user $u_i$, given the $n-1$ context information $\mathcal{O}_i$, our task is to build a SRS model $M$ (i.e., HPM) to learn the user preference dynamics from the $\mathcal{O}_i$ and then generate the recommended list which can best satisfy the user’s preference at the moment $t_n$.

\subsection{Framework Overview}

\begin{figure*}[htp]
	\centering
	\includegraphics[width=1\linewidth]{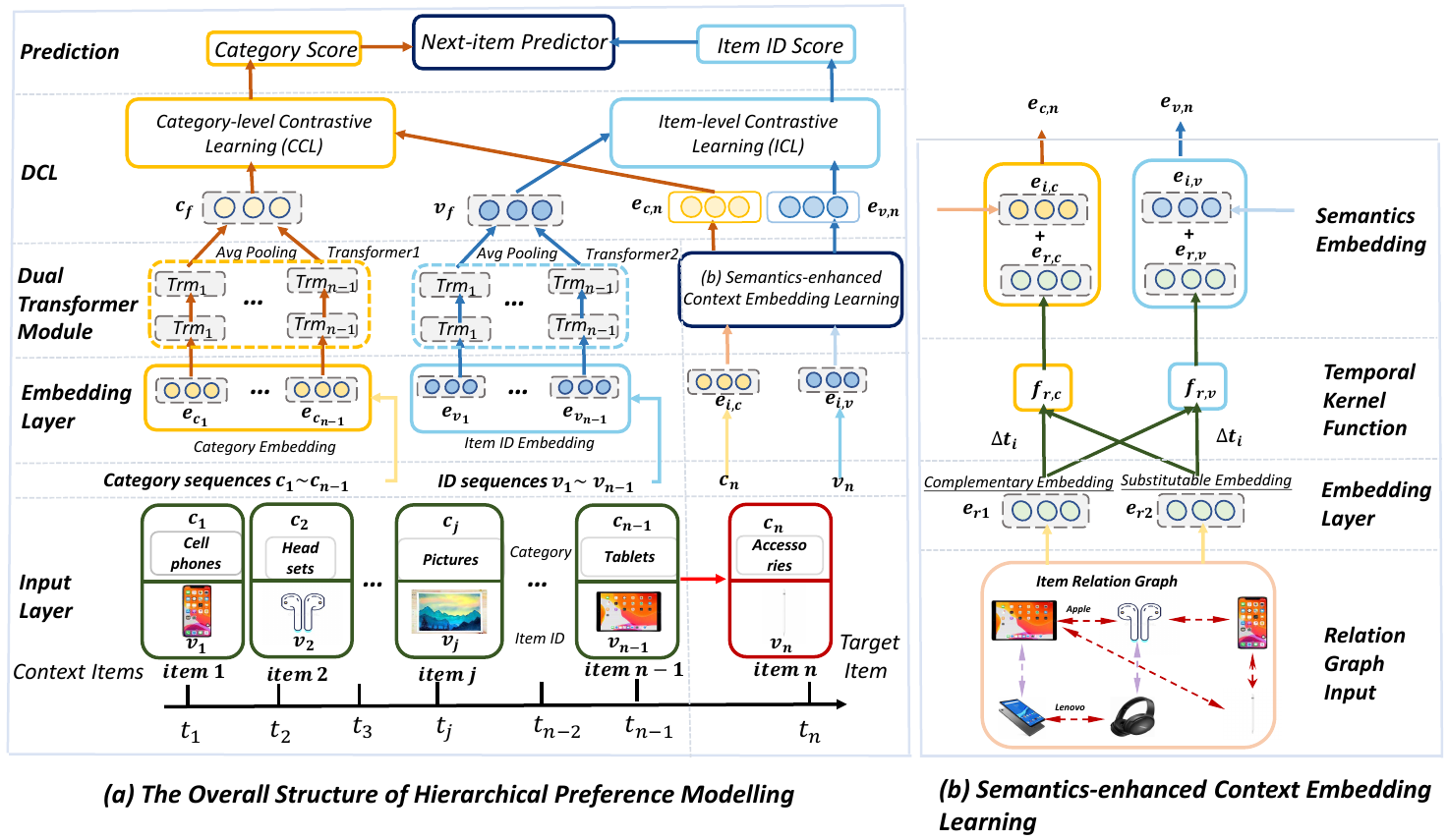}
	\caption{(a)The overall framework of our Hierarchical Preference modeling (HPM), which is composed of three main components: Dual Transformer module for hierarchical preference modeling, Semantics-enhanced Context Embedding Learning (SCEL), and Dual-Contrastive Learning (DCL) scheme;  (b) Semantics-enhanced Context Embedding Learning (SCEL) leverages the relations between context items and target items to enhance the representation of target item embedding.}
	\label{fig:framework}
\end{figure*}

The framework for our proposed HPM is shown in Figure \ref{fig:framework}, which is composed of three main components: (1) Dual Transformer (DT) for hierarchical preference modeling, (2) Semantics-enhanced Context Embedding Learning (SCEL), and (3) Dual hierarchical preference Contrastive Learning (DCL). 

\subsection{Embedding Layers}
\label{ISM}

We use two levels of embedding (item ID embedding and category-type embedding) as the input to our model to better model preference dynamics throughout users' interaction history. Also, we use the classical TransE~\cite{TransE} method to pre-train the relationships between items to enhance the semantic relevance of items the user purchased. We use $E$ to represent all the embedding set, where $e_v \in E_V$ denotes the item ID embedding, $e_r \in E_R$ denotes the relation embedding and $e_c \in E_C$ denotes the item category embedding.

\textbf{Item ID Embedding.} The input sequence is made up of item IDs. To obtain a unique dense embedding for each item ID, we use a linear embedding layer. For the user, the ID representation $e_v$ of the item transitions relatively sharply, and this representation is suitable for capturing rapid changes in user short-term preference.

\textbf{Category Type Embedding.} 
Similar to item ID embedding, we use a linear embedding layer to represent category features $e_c$. For users, the high-level category preference changes relatively smoothly. It is closer to the stable preference of the user.

\textbf{Knowledge Graph Embedding.} Meanwhile, we introduce knowledge graph embedding to enhance the correlation among items by modeling the direct semantic relationships between features of items that users interact with.
Without losing generality, we follow the previous work \cite{chorus} and use \textit{TransE} \cite{TransE} to pre-train item and relation embeddings: 
\begin{equation}
    f(v_h, r, v_t)=\|\mathbf{e_{v,h}}+\mathbf{e_r}-\mathbf{e_{v,t}}\|_{2}^{2}, \
       \hspace{1ex}
    f(c_h, r, c_t)=\|\mathbf{e_{c,h}}+\mathbf{e_r}-\mathbf{e_{c,t}}\|_{2}^{2},
\end{equation}
where $f(\cdot)$ denotes the loss function of  TransE. $v_h$ and $v_t$ denote the IDs of head item and tail item respectively; $c_h$ and $c_t$ denote the category of items $v_h$ and $v_t$ respectively while $r \in \mathcal{R}$ denotes the relation ID between items $v_h$ and $v_t$. 

\subsection{Dual Transformer for Hierarchical Preference Modeling \label{preference}}

To accurately model the hierarchical preference dynamics, we take the state-of-the-art sequential recommendation model, SASRec~\cite{SARS}, as the base architecture of our proposed HPM model. SASRec uses the self-attention transformer to capture users’ dynamic preference over time. However, the existing SASRec model can only model users' preferences towards specific items at the item level based on the ID information of items in sequences. It cannot model the coarse-grained user preference at a higher level (e.g., item category level), which are relatively stable and changes slowly compared with the users' fine-grained preferences towards items at the item level. To well capture both the high-level user preference at the category aspect and the low-level user preference shift at the item aspect for better characterizing a user for the next-item recommendation, we develop a novel Dual-Transformer (DT) module to equip the HPM framework. DT module takes item ID $V_i$ and item category $C_i$ as the input of every single transformer respectively. 

We first obtain the pre-trained item embedding $E_V \in \mathbf{R}^{|V| \times d}$ and category embedding $E_C \in \mathbf{R}^{|C| \times d}$ in Section \ref{ISM}. Here $|V|$ and $|C|$ denote the number of items and the number of categories in the dataset respectively, and $d$ denotes the dimension of item embedding and category embedding. To comprehensively capture the order information over items in sequences, we introduce the position embedding matrix $E_P \in \mathbf{R}^{L \times d}$ where $L$ is the length of interaction sequences and $d$ is the embedding dimension. Based on these learnable embedding matrices, we can obtain the embedding vector $e_v = E_{V}(v_i) \in \mathbf{R}^{1 x d}$ of a given item $v_i$, its category embedding $e_c = E_{C}(c_i) \in \mathbf{R}^{1 x d}$, and the embedding $p_i = E_{P}(p_i) \in \mathbf{R}^{1 x d}$ of the item position in the sequence. 
Therefore, given a user-item interaction sequence $\mathcal{S}_i$ of user $u_i$, the position-sensitive representation of an item $v_i \in \mathcal{S}_i$ and the representation of its corresponding category $c_i$ are calculated as: 
\begin{equation}
   e_{i,v} = e_{i,v} + p_i, \hspace{2ex} 
   e_{i,c} = e_{i,c} + p_i,
\end{equation}
Once the item and category representations are ready, we choose the self-attention mechanism to well capture the relationships between each interacted item in the sequence $\mathcal{S}_i$ and their context information in $\mathcal{S}_i$.
Specifically, we choose multi-head self-attention to model the $n-1$ historical items interacted by user $u_i$ to learn the complex relationships between each item and its corresponding contextual items in the sequence. Multi-head self-attention uses different linear projection functions to map the history interaction embedding into $h$ different sub-spaces so as to obtain richer information from different subspaces. 

After applying the self-attention mechanism to each head, we first concatenate and then project the concatenated multi-head embedding back to the same dimension of $e_{i,v}$. Specifically, the multi-head attention model is formulated below:  
\begin{equation}
\mathrm{MultiHead}(H_{i,*})= W^O \text{concat}(\text{head}_1;\text{head}_2;\cdots;\text{head}_h), \\
\end{equation}
\begin{equation}
\operatorname{head}_i =\text{Attention}\left(H_{i,*} W_i^Q, H_{i,*} W_i^K, H_{i,*} W_i^V\right),
\end{equation}
\begin{equation}
\mathrm{A}_{i} = \operatorname{Attention}(Q, K, V)=\operatorname{softmax}\left(\frac{Q K^{\top}}{\sqrt{d / h}}\right)V, 
\end{equation}
where $H_{i,*}$ denotes all the context embeddings in the historical interactions. $*$ means the input can be item ID embedding $e_v$ sequence or item category embedding $e_c$ sequence. $W_i^Q, W_i^K, W_i^V \in R^{d x \frac{d}{h}} $ are the linear transformation weight matrices of query, key, and value of the self-attention model respectively. $W_O \in R^{d x d} $ is the transformation weight matrix of output. $h$ denotes the number of heads. After the linear projection, we add the position-wise feed-forward network to introduce the non-linearity into our model to make it more powerful for capturing complex non-linear relations. Meanwhile, following the successful practice in previous work \cite{CL4Rec}, we introduce the $LayerNorm$, $Dropout$, and residual connection modules to reduce the over-fitting issue. The mathematical formulations of these layers are given below: 
\begin{equation}
FFN(\mathrm{A}_{i})=\operatorname{ReLU}(\mathrm{A}_{i}W_{1}+b_{1})W_{2}+b_{2},
\end{equation}
\begin{equation}
    H_{i,*} = LayerNorm(H_{i,*} + Dropout(FFN(\mathrm{A}_i))), 
\end{equation}
where $FFN$ represents a fully connected feed-forward network and $LayerNorm$ denotes a normalization layer. $H_{i,*}$ represents either $e_v$ or $e_c$ in the user sequence $\mathcal{S}_i$. For high- and low-level user preference modeling, we have:
\begin{equation}
v_f=\frac{1}{L} \sum_{l=1}^{L} S_{i,v_l},
\hspace{2ex}
c_f=\frac{1}{L} \sum_{l=1}^{L} S_{i,c_l},
\end{equation}
where $l$ means the $l$-th position in sequence $S_i$. We use the average pooling operation to get the user item-level historical representation $v_f$ and user category-level representation $c_f$ respectively, which comprehensively model both the low-level preference dynamics and high-level preference dynamics.

\subsection{Semantics-enhanced Context Embedding Learning}
\label{SCEL}

In the section, we will introduce our \textbf{S}emantic-enhanced \textbf{C}ontext \textbf{E}mbedding \textbf{L}earning (SCEL) module.
Although the multi-granular preference dynamics modeling can handle different-level user preference dynamics, there are many cases that only contain sparse interactions in the short sequences. Hence, we propose to leverage the context relation information to enhance the representation ability of short user sequences.
Specially, our SCEL leverages the explicit relationship to enhance the connection between history items and the target item. We mainly focus on the two different relations: also\_buy and also\_view.
However, we believe that the intensity of the relationship among items is not fixed, but changes over time. Thus, we need to introduce different temporal time functions to model this kind of fluctuation.  

For also\_buy relations, we regard them as complementary relations. For example, when a user bought an iPad, he has a very high probability to buy an Apple pencil in the short term, which is the bundle product of the iPad and shares the same brand with the iPad. However, the probability of buying an Apple pencil decreases as time goes by since the user preference shifts to another field. Thus, it is suitable for us to choose a normal distribution to model this user preference decay process:
\begin{equation}
    \mathcal{\phi}^1_v(\Delta t) = N( \Delta t | 0, \sigma_v),
   \hspace{2ex}
    \mathcal{\phi}^1_c(\Delta t) = N( \Delta t | 0, \sigma_c),
\end{equation}
where $\Delta t$ denotes the time interval between the history item and the target item. $N$ denotes the normal distribution. 0 denotes the $\mu$ = 0 and $\sigma_v$, $\sigma_c$ denote the item id and category variance respectively. $\sigma_v$ curves the corresponding item-level relation decay while $\sigma_c$ curves the corresponding item-level relation decay, which can map to the previous hierarchical
preference modeling. 

For also\_view, we regard them as substitute relations. For instance, if a user has bought an iPhone recently, he/she is unlikely to buy the same type of product in the near future. But the probability would increase with time moves on, similar items would also need to be replaced. User perhaps buys similar products in the long term. Thus, we combine short-term negative- and long-term positive temporal kernel functions.
\begin{equation}
    \mathcal{\phi}^2_v(\Delta t) = -N( \Delta t | 0, \sigma_v) + N( \Delta t| \mu_v, \sigma_v),
\end{equation}
\begin{equation}
    \mathcal{\phi}^2_c(\Delta t) = -N( \Delta t | 0, \sigma_c) + N( \Delta t| \mu_c, \sigma_c),
\end{equation}
Then, incorporating both item and category temporal dynamics of different relations, we can obtain the \textbf{semantics-enhanced context embedding} of target item:
\begin{equation}
    e_{v,n} = e_{v,n} + e_{r,v}, e_{r,v} = \sum_{r \in \mathcal{R}}{f_r(\mathcal{S}_{i,v},t,e_v) \cdot e_r},
\end{equation}
\begin{equation}
    e_{c,n} = e_{c,n} + e_{r,c},  e_{r,c} = \sum_{r \in \mathcal{R}}{f_r(\mathcal{S}_{i,c},t,e_i) \cdot e_r}
\end{equation}
\begin{equation}
   f_{r1}(\mathcal{S}_{i,v}, v, t) = \sum_{v',t'}{\mathrm{I}_r(v,v') \cdot \phi(t-t') },
\end{equation}
\begin{equation}
   f_{r2}(\mathcal{S}_{i,c}, c, t) = \sum_{c',t'}{\mathrm{I}_r(c,c') \cdot \phi(t-t') },
\end{equation}
where $e_{v,n}, e_{c,n}$ denote the pre-trained target item embedding and category embedding, $r$ denotes the different item relations between history item $\mathcal{S}_{i,*}$ ($*$ means both item ID $v$ and category $c$) and target item $e_{v,*}$. $e_r$ denotes the embedding of relation. $I_r$ denotes the indicator function, if history item has explicit relation with target item, $I(i,i')=1$. $I(i,i')=0$ vice versa. We can treat the association between history items and target item as a kind of multi-excitation, which means the user multi-granularity history preference impact on current user preference with dynamics. Since users' interaction behaviors always occur the relation-oriented patterns, the semantics-enhanced context can better capture and incorporate the hidden semantic relations between items to generate more informative sequence context embedding for next-item prediction, especially in sparse interaction situations. 

\subsection{Dual Hierarchical Preference Contrastive Learning \label{contrast}}
Currently, contrastive learning tends to solve the data sparsity problem of sequential recommendation by providing additional supervised signals by augmenting the original sequences~\cite{ContraRec,S3-Rec, CL4Rec,ICL}. 
However, current CL-based SRSs only involve a single contrast
based on the low-level preference indicated by item ID information,
overlooking the contrast built on the high-level preference indicated by
item category information, which results in the lack of users' high-level preference modeling and future demand may not be substantially learned, especially on highly sparse datasets~\cite{CLproblem2}. Also, the absence of category-level contrastive learning might cause the loss of some important constraint signal to connect (resp. distinguish) items from the same (resp. different) categories, further impeding the recommendation performance.

Besides, current contrastive learning in RS adopts maximizing the mutual information between different augmented views from the original user sequence to enhance the model performance. adopting augmentation such as reordering, random masking, and dropout would disrupt intrinsic patterns within the raw data. Especially for the personalized sequential recommendation, these approaches might perturb the temporal relationship among sequential items for each user, which is unsuitable for our SCEL module. 

Furthermore, as aforementioned in the hierarchical preference dynamics modeling, we observe that in most of the user's interaction history sequence, the user's preference for item categories is more stable, and the interaction changes for actual single items are more drastic. So for better preference dynamics modeling, it is more suitable to adopt dual contrastive learning to model the user hierarchical preference dynamics. 
To be more specific, we assume that users have relatively stable and slow-changed high-level (i.e. category) preferences. Besides, they have relatively drastic and fast-changed low-level (i.e. item-ID) preferences. Both of them are influenced by the aforementioned context information (Section \ref{SCEL}). Then based on this assumption, we propose the Dual-Contrastive Learning (DCL) module (i.e. category-level and item-level contrastive learning) in terms of user high- and low-level preference, which explicitly considers the dynamic changes in user preference.
To be specific, we utilize the history item embedding $S_{i,v}$ and $S_{i,c}$ as hierarchical user representation, semantics-enhanced context embedding of target item as a positive sample, other target items in the same batch as negative samples. 
\begin{equation}
    \mathcal{L}_{cl_{item}}(e_{v,n}, v_f) =  -\log \frac{\exp(\operatorname{sim}(e_{v,n}, v_f))}{\exp(\operatorname{sim}(e_{v,n}, v_f))) + \sum_{v_f^- \in V_f}{\operatorname{sim}(e_{v,n}, v_f^-)}   },
\end{equation}
\begin{equation}
    \mathcal{L}_{cl_{cate}}(e_{c,n}, c_f) =  -\log \frac{\exp(\operatorname{sim}(e_{c,n}, c_f))}{\exp(\operatorname{sim}(e_{c,n}, c_f))) + \sum_{c_f^- \in C_f}{\operatorname{sim}(e_{c,n}, c_f^-)}   },
\end{equation}
$e_{v,n}$ denotes historical item ID representation and $e_{c,n}$ denotes historical category type representation. 
$e_f$ denotes the semantics-enhanced context-customized embedding of the target item ID and $e_c$ denotes the semantics-enhanced context-customized embedding of the target category. $\operatorname{sim}(\cdot)$ means the distance function, we choose the cosine function to calculate the similarity between context embedding and target item embedding. Thus, $\mathcal{L}_{cl_{item}}$ enhances the item-level context-customized embedding learning for the low-level user preference learning and 
$\mathcal{L}_{cl_{cate}}$ enhances the category-level context-customized embedding learning for the high-level user preference learning.  
Combining with item- and category-level CL, we get the final Dual Contrastive Learning (DCL) loss function:
\begin{equation}
    \mathcal{L}_{cl} = \mathcal{L}_{cl_{item}} + \mathcal{L}_{cl_{cate}}.
\end{equation}
\subsection{Training and Optimization}

To learn the parameters of our HPM in the sequential recommendation, we adopt the pairwise ranking loss (BPR loss) to optimize our model: 
\begin{equation}
    \mathcal{L}_{rec}=-\sum_{u \in \mathcal{U}} \sum_{i=1}^{N_{u}} \log \sigma\left(\hat{y}_{u i}-\hat{y}_{u j}\right),
\end{equation}
\begin{equation}
    \hat{y}_{u i} = e_{v,n}^Tv_{f,i} + e_{c,n}^Tc_{f,i}, \hspace{2ex}  
\hat{y}_{u j} = e_{v,n}^Tv_{f,j} + e_{c,n}^Tc_{f,j},
\end{equation}
where $\sigma(\cdot)$ represents the sigmoid function, $\hat{y}_{u i}$ represents the preference score of user $u$ to positive item $i$ while $\hat{y}_{u j}$ represents the preference score of user $u$ to negative item $j$. 
Consequently, those top-k items with high possibilities are selected according to $\hat{y}$ to form the recommendation list. 
We adopt multi-task learning optimizing the ranking loss and contrastive loss jointly. The joint loss is as follows:
\begin{equation}
    \mathcal{L}_{joint} = \mathcal{L}_{rec} + \lambda \mathcal{L}_{cl}.
\end{equation}
where $\lambda$ means the CL loss coefficient, it controls the strength of DCL.
Our model is implemented using Pytorch 1.10 running under Python 3.6.8 environment. Model parameters are learned by minimizing the total loss $\mathcal{L}_{joint}$ based on a mini-batch learning procedure. More model training settings are discussed in Section \ref{Parameter Settings}.
Our experiments are conducted with a single NVIDIA TITAN RTX GPU with 24 GB RAM.

\section{Experiments}

\subsection{Data Preparation}

We conduct extensive experiments on the public real-world Amazon dataset \cite{AmazonDataset}, which has been commonly used for sequential recommendations~\cite{FusingM,rakkappan2019context}.
Specifically, we choose six representative sub-datasets from Amazon dataset, which correspond to six top-level product categories respectively: \textit{Grocery and Gourmet Food} (denoted as Grocery), \textit{Sports and Outdoors} (Sports),  \textit{Beauty}, \textit{Clothing Shoes and Jewelry} (Clothing), \textit{Cellphones and Accessories} (Cellphones) and \textit{Toys and Games} (Toys). We follow the commonly followed practice~\cite{SLRS, KDA, chorus, CL4Rec} in ReChorus experiment~\footnote{https://github.com/THUwangcy/ReChorus} to prepare for the experimental data and build training-test instances. 
Following the previous work~\cite{chorus, SLRS}, we take the 'also buy' as complementary and 'also view' as substitute relations.
We further introduce two extra item relations, namely 'same brand' as complementary and 'same category with similar price' as substitute relations. 

We follow the common practice in handling sequential recommendation datasets \cite{SLRS,chorus,KDA}. In detail, we only keep the ‘5-core’ datasets, in which all users and items have at least 5 interactions. We set the maximum interaction history $len(\mathcal{S}_u)$ as 20. If the $len(\mathcal{S}_u)$ is more than 20, we adopt the latest 20 interactions; otherwise, we pad them with 0 to make up to 20 interactions. Finally, we adopt the leave-one-out evaluation by following previous works~\cite{ACF,KabburNK13,KDA}. To be specific, we choose the most recent interacted item as the test target item and the second last item as the validation target item. 

\subsection{Experimental Setting \label{setting}}

\renewcommand\arraystretch{1.1}
\begin{table*}[!h]
    \caption{Overall performance. Bold scores represent the highest results of all methods. Underlined scores stand for the highest results from previous methods. We perform 5 times experiments and report the average result and our model achieves the state-of-the-art result among all baseline models. $^*$ denotes the improvement is significant at p < 0.05.}
    \small
    \setlength\tabcolsep{2pt}  
    \begin{tabular}{c|l|ccccccccccccc|c|c}
    \toprule
         Dataset& Metric      & FPMC & GRU4Rec & Caser & SASRec & TiSASRec & SLRS+ &Chorus &DIF &CL4Rec &S3Rec &ContraRec &DuoRec &KDA & HPM & Improv.\\
         \midrule        
         \multirow{8}*{Beauty}&HR@5&0.3392&0.3202&0.3210&0.3666&0.3872&0.4339&0.4536 &0.4102 &0.3754&0.3812&0.4012&0.4123 &\underline{0.4921}&$\textbf{0.5141}^*$&4.78\%\\
                              &HR@10&0.4290&0.4311&0.4345&0.4590&0.4559&0.5337&0.5698&0.5209   &0.4660&0.4810&0.4962&0.5039 &\underline{0.6076}&$\textbf{0.6298}^*$&3.65\%\\
                              &HR@20&0.5393&0.5693&0.5757&0.5743&0.5700&0.6361&0.6838&0.6421   &0.5830&0.6057&0.6065& 0.6131 &\underline{0.7221}&$\textbf{0.7424}^*$&2.81\%\\
                              &HR@50&0.7511&0.7973&0.8097&0.7756&0.7745&0.8033&0.8536&0.8284   &0.8013&0.8146&0.8530& 0.8033 & \underline{0.8853}&$\textbf{0.8961}^*$&1.22\%\\

                              &NDCG@5&0.2558&0.2271&0.2246&0.2797&0.2904&0.3319&0.3386&0.3016    &0.2842&0.3073&0.3406& 0.3158 & \underline{0.3666}&$\textbf{0.3864}^*$&5.40\%\\
                              &NDCG@10&0.2848&0.2628&0.2612&0.3094& 0.3036&0.3642&0.3762&0.3374  &0.3134&0.3379&0.3784&0.3454 & \underline{0.4040}&$\textbf{0.4239}^*$&4.93\%\\
                              &NDCG@20&0.3125&0.2976&0.2967&0.3385& 0.3324&0.3900&0.4050&0.3679  &0.3429&0.3657&0.4058&0.3729 & \underline{0.4329}&$\textbf{0.4524}^*$&4.50\%\\
                              &NDCG@50&0.3542&0.3426&0.3430&0.3782& 0.3728&0.4232&0.4386&0.4048  &0.3859&0.4041&0.4397&0.4104 &  \underline{0.4653}&$\textbf{0.4830}^*$&3.80\%\\
         \midrule        
         \multirow{8}*{Clothing}&HR@5&0.2020&0.2142&0.2269&0.2301&0.2722& 0.3029&0.3826&0.2977     &0.2600&0.2787  &0.3798&0.2781&  \underline{0.3863}&$\textbf{0.4526}^*$&17.16\%\\
                                &HR@10&0.2834&0.3142&0.3354&0.3571&0.3808& 0.3904&0.4916&0.4068     &0.3693&0.3797 &0.4891&0.3799 & \underline{0.4991}&$\textbf{0.5748}^*$&15.17\%\\
                                &HR@20&0.4014&0.4517&0.4892&0.5097&0.5142&0.5004&0.6141&0.5403      &0.5161&0.5166 &0.6094& 0.5155 & \underline{0.6270}&$\textbf{0.7064}^*$&12.66\%\\
                                &HR@50&0.6553&0.7143&0.7531&0.7453&0.7405&0.6948&0.8046&0.7600      &0.7839&0.7665 &0.8028& 0.7650 & \underline{0.8317}&$\textbf{0.8803}^*$&5.84\%\\
                                
                                &NDCG@5&0.1442&0.1461&0.1548&0.1642&0.1927&0.2329&0.2840&0.2130   &0.1854&0.2016 &0.2840&0.2012 &  \underline{0.2880}&$\textbf{0.3387}^*$&17.61\%\\
                                &NDCG@10&0.1703&0.1783&0.1897&0.1946&0.2278&0.2611&0.3192&0.2481   &0.2206&0.2341  &0.3193&0.2339&      \underline{0.3244}&$\textbf{0.3781}^*$&16.55\%\\
                                &NDCG@20&0.2000&0.2130&0.2284&0.2349&0.2613& 0.2888&0.3501&0.2817  &0.2576&0.2686 &0.3496&0.2680&        \underline{0.3567}&$\textbf{0.4114}^*$&15.34\%\\
                                &NDCG@50&0.2499&0.2647&0.2807&0.2923&0.3060& 0.3271&0.3878&0.3252   &0.3103&0.3179 &0.3879&0.3172&  \underline{0.3973}&$\textbf{0.4460}^*$&12.26\%\\
         \midrule
         \multirow{8}*{Sports}&HR@5&0.3260&0.3015&0.3145&0.3414&0.3475& 0.3900&0.4544&0.3945     &0.3719&0.3960 &0.4544& 0.3948 &  \underline{0.4672}&$\textbf{0.4984}^*$&6.68\%\\
                              &HR@10&0.4373&0.4301&0.4423&0.4566&0.4608& 0.4827&0.5823&0.5197    &0.5035&0.5160   &0.5823&0.5151&        \underline{0.6021}&$\textbf{0.6306}^*$&4.73\%\\
                              &HR@20&0.5748&0.5918&0.6039&0.5943&0.6003& 0.5961&0.7162&0.6612    &0.6555&0.6567  &0.7162& 0.6553 &      \underline{0.7392}&$\textbf{0.7638}^*$&3.33\%\\
                              &HR@50&0.8070&0.8412&0.8496&0.8096&0.8131& 0.7784&0.8855&0.8575    &0.8652&0.8619    &0.8855& 0.8631&       \underline{0.9042}&$\textbf{0.9198}^*$&1.72\%\\
                              
                              &NDCG@5&0.2381&0.2085&0.2175&0.2494&0.2535& 0.3013&0.3354&0.2852    &0.2681&0.2906   &0.3354&0.2894& \underline{0.3402}&$\textbf{0.3708}^*$&8.25\%\\
                              &NDCG@10&0.2740&0.2498&0.2588&0.2866&0.2901&0.3311&0.3767&0.3257    &0.3106&0.3294  &0.3767& 0.3282& \underline{0.3838}&$\textbf{0.4136}^*$&8.99\%\\
                              &NDCG@20&0.3086&0.2905&0.2995&0.3214&0.3253& 0.3597&0.4106&0.3615   &0.3489&0.3648 &0.4106& 0.3636&  \underline{0.4185}&$\textbf{0.4474}^*$&6.91\%\\
                              &NDCG@50&0.3546&0.3400&0.3484&0.3641&0.3675& 0.3957&0.4443&0.4005    &0.3907&0.4056   &0.4443&0.4049& \underline{0.4515}&$\textbf{0.4785}^*$&5.98\%\\
                              
        \midrule      
        \multirow{8}*{Cellphone}&HR@5&0.4003&0.3015&0.3937&0.4439&0.4520& 0.4696&0.4697&0.4718    &0.4085&0.4505 &0.4829&  0.4745&      \underline{0.5497}&$\textbf{0.5835}^*$&6.15\%\\
                              &HR@10&0.5098&0.4301&0.5309&0.5595&0.5767& 0.5641&0.5929& 0.5951     &0.5415&0.5819   &0.5994& 0.5920&       \underline{0.6745}&$\textbf{0.7050}^*$&4.52\%\\
                              &HR@20&0.6321&0.5918&0.6810&0.6817&0.7022& 0.6637&0.7152& 0.7157     &0.6861&0.7147  &0.7211&  0.7151&      \underline{0.7923}&$\textbf{0.8225}^*$&3.81\%\\
                              &HR@50&0.8277&0.8412&0.8849&0.8676&0.8708& 0.8172&0.8695& 0.8749     &0.8825&0.8880    &0.8831& 0.8792&      \underline{0.9263}&$\textbf{0.9428}^*$&1.78\%\\
                              
                              &NDCG@5&0.3027&0.2085&0.2800&0.3353&0.3344& 0.3634&0.3530&0.3526   &0.2967&0.3287   &0.3673& 0.3602&       \underline{0.4119}&$\textbf{0.4487}^*$&8.76\%\\
                              &NDCG@10&0.3381&0.2498&0.3243&0.3727&0.3748&0.3939&0.3929&0.3925    &0.3396&0.3712  &0.4050& 0.3983&       \underline{0.4523}&$\textbf{0.4882}^*$&7.94\%\\
                              &NDCG@20&0.3690&0.2905&0.3622&0.4036&0.4065& 0.4191&0.4238&0.4230    &0.3761&0.4047 &0.4358&  0.4294&       \underline{0.4821}&$\textbf{0.5179}^*$&7.43\%\\
                              &NDCG@50&0.4077&0.3400&0.4028&0.4370&0.4401& 0.4495&0.4545& 0.4548   &0.4152&0.4393   &0.4681& 0.4620&       \underline{0.5089}&$\textbf{0.5419}^*$&6.48\%\\

        \midrule       
        \multirow{8}*{Toys}   &HR@5&0.3373&0.2902&0.2898&0.3602&0.3475& 0.4368&0.4124&0.3843     &0.3627&0.3759 &0.4015& 0.4001&       \underline{0.4805}&$\textbf{0.4927}^*$&2.54\%\\
                              &HR@10&0.4233&0.4060&0.4103&0.4570&0.4608& 0.5345&0.5203&0.4924    &0.4643&0.4731   &0.4958&0.4953&        \underline{0.5882}&$\textbf{0.6039}^*$&2.67\%\\
                              &HR@20&0.5283&0.5546&0.5590&0.5700&0.6003& 0.6440&0.6443&0.6149     &0.5900&0.5972  &0.6181& 0.6164&      \underline{0.7019}&$\textbf{0.7211}^*$&2.74\%\\
                              &HR@50&0.7482&0.8067&0.8107&0.7789&0.8131& 0.8012&0.8277&0.8178     &0.8208&0.8101    &0.8256& 0.8244 &      \underline{0.8772}&$\textbf{0.8922}^*$&1.71\%\\
                              
                              &NDCG@5&0.2583&0.1974&0.1947&0.2738&0.2535& 0.3490&0.3132&0.2829    &0.2630&0.2811   &0.3067&0.3046&        \underline{0.3660}&$\textbf{0.3807}^*$&4.02\%\\
                              &NDCG@10&0.2860&0.2348&0.2336&0.3050&0.2901&0.3804&0.3480&0.3178    &0.2957&0.3124  &0.3371&0.3355&        \underline{0.4007}&$\textbf{0.4166}^*$&3.97\%\\
                              &NDCG@20&0.3124&0.2721&0.2710&0.3334&0.3253& 0.4081&0.3793&0.3488    &0.3273&0.3437 &0.3679&0.3660 &        \underline{0.4294}&$\textbf{0.4462}^*$&3.91\%\\
                              &NDCG@50&0.3556&0.3220&0.3207&0.3747&0.3675& 0.4392&0.4156&0.3890    &0.3707&0.3858   &0.4089&0.4071&         \underline{0.4642}&$\textbf{0.4803}^*$&3.47\%\\

        \midrule        
        \multirow{8}*{Grocery}&HR@5&0.3618&0.3737&0.3145&0.3925&0.4069& 0.4378&0.4513&0.4301    &0.3669&0.4029 &0.4268&0.4269& \underline{0.5168}&$\textbf{0.5432}^*$&5.11\%\\
                              &HR@10&0.4419&0.4793&0.4423&0.4801&0.5232& 0.5523&0.5818&0.5376    &0.4624&0.5051&0.5132&0.5127& \underline{0.6314}&$\textbf{0.6476}^*$&2.57\%\\
                              &HR@20&0.5432&0.6013&0.6039&0.5822&0.6350& 0.6517&0.6956&0.6465    &0.5737&0.6260&0.6170&0.6213& \underline{0.7401}&$\textbf{0.7514}^*$&1.53\%\\
                              &HR@50&0.7511&0.8245&0.8496&0.7709&0.8217& 0.7995&0.8576&0.8273     &0.7964&0.8202&0.8157&0.8190& \underline{0.8901}&$\textbf{0.8999}^*$&1.10\%\\
                              
                              &NDCG@5&0.2816&0.2684&0.2175&0.2941&0.2906& 0.3266&0.3223&0.3122   &0.2702&0.2969 &0.3291&0.3293& \underline{0.3892}&$\textbf{0.4088}^*$&5.04\%\\
                              &NDCG@10&0.3073&0.3024&0.2588&0.3231&0.3283&0.3637&0.3647&0.3470   &0.2992&0.3299 &0.3571&0.3570& \underline{0.4264}&$\textbf{0.4428}^*$&3.85\%\\
                              &NDCG@20&0.3328&0.3331&0.2995&0.3488&0.3565& 0.3888&0.3934&0.3745   &0.3286&0.3604&0.3831&0.3844& \underline{0.4539}&$\textbf{0.4689}^*$&3.30\%\\
                              &NDCG@50&0.3737&0.3772&0.3484&0.3861&0.3934& 0.4180&0.4256&0.4104      &0.3745&0.3988&0.4224&0.4235&  \underline{0.4838}&$\textbf{0.4985}^*$&3.04\%\\
         \bottomrule
    \end{tabular}
    \label{tab:overall}
\end{table*}

\subsubsection{\textbf{Baselines for Comparisons}}
        Our task is essentially the next item prediction in sequence recommendation~\cite{GRU4Rec1,IJCAISurvey,FPMC}. 
Hence, we carefully select 13 representative and/or state-of-the-art approaches from different classes for sequential recommendation as baselines. 
\textbf{1) Traditional sequential recommendation methods}: \textit{FPMC}~\cite{FPMC}: This model is based on personalized transition graphs over underlying Markov chains and combines the matrix factorization for sequential recommendation. \textit{GRU4Rec}~\cite{GRU4Rec1}: This model uses the GRU to model the user interaction sequence for recommendation.
\textit{Caser}~\cite{Caser}: This model embeds items in user interaction history as images by using convolutional filters for recommendation. 
\textit{SASRec}~\cite{SARS}: This model leverages users’ longer-term
semantics as well as their recent actions simultaneously for the accurate next-item recommendation. \textbf{2) Temporal sequential recommendation methods}: \textit{TiSASRec}~\cite{TiSASRec}: This model leverages the timestamp and time intervals between user-item interactions for the next item prediction.  
\textit{SLRS+} \cite{SLRS}: SLRS combines Hawkes process and MF into one framework for modeling the user repeat consumption in sequential recommendation. Since the Amazon dataset removes the repeat consumption in the test set, SLRS+ uses the Hawkes process to model the relations including also view and also buy behaviors.
\textit{Chorus} \cite{chorus}: This model is a state-of-the-art method by considering item relations and temporal evolution.
\textit{KDA} \cite{KDA}: KDA devises relational intensity and frequency-domain embeddings to adaptively determine the importance of historical interactions. \textbf{3) Category-aware sequential recommendation methods}: \textit{DIF} \cite{XieZK22}: DIF-SR diverts the side information into the attention layer and decouples the attention calculation of various side information and item representation. 
\textbf{4) Contrastive sequential recommendation methods}:
\textit{ContraRec}: a novel context-context contrastive signals learning method for sequential recommendation ~\cite{ContraRec}.
\textit{CL4SRec}: This model utilizes the contrastive learning between augmented historical sequences and original historical sequences~\cite{CL4Rec}.
\textit{S3Rec:} It designs four pretext tasks for context-aware recommendation and then finetunes on the next-item recommendation task, which is a state-of-the-art method based on self-supervised learning~\cite{S3-Rec}. 
\textit{DuoRec:} A CL-based SR model that utilizes both the feature-level dropout masking and the supervised positive sampling to construct contrastive samples~\cite{DuoRec}.
\subsubsection{\textbf{Evaluation Metric}}
We adopt the commonly used NDCG@K and Hit Rate@K to evaluate our model~\cite{SARS, chorus}. 1) \textit{NDCG@K}: a position-aware ranking metric that takes the normalized value of discounted cumulative gain. 2) \textit{Hit Rate@K}: the fraction of times that the ground-truth next item is among the top K items. Both of them are applied with K chosen from \{5,10,20,50\}. 
We evaluate the ranking results with 99 randomly selected negative items following previous works~\cite{SLRS,chorus,KDA}. Besides, a paired t-test with p<0.05 is used for the significance test by following ~\cite{WangRMCMR19,WangXZWS22}.

\subsubsection{\textbf{Parameter Settings}}
\label{Parameter Settings}
For fair comparisons and constrained by limited computing resources, we set the embedding size and batch size as 64 for all the models. All the other model parameters including hyper-parameters of both baseline methods and our method are well-tuned in the same way on the validation set. In the training process, we follow the previous setting \cite{SLRS}\cite{chorus}\cite{KDA}, setting the number of the negative sample as 1.
For multi-head self-attention-based methods, the number of heads and layers are tuned in \{1,2,3,4\} and \{1,2,3\} respectively.
The maximum number of training epochs for all the datasets is set to 200. If the model's performance on the validation set decreases for 10 consecutive rounds, the training will early stop. For our model, we set the self-attention layer as 1 and attention heads as 4, the dimension of the embedding as 64, and $\lambda$ as 1 after tuning, which are discussed carefully during the experimental result analysis in Section \ref{PST}.
Then our model is optimized by an Adam with a learning rate of 1e-5 for item knowledge embedding pre-training and 1e-6 for the main model training. 

To achieve the best performance, we carefully tune the hyper-parameters for all the baselines according to the result on the validation set. In Caser, the number of horizon convolution kernels is set as 64, the number of vertical convolution kernels is set as 32 and the union window size is set as 5. In GRU4Rec, the dimension of the hidden layer is set as 64. In SASRec and TiSARS, the number of heads is set as 1. In SLRS+ and Chorus, the learning rate is set as 5e-4. In KDA, the number of heads is set as 4 and the learning rate is set as 1e-3. 

\subsection{Overall Performance Comparison}
\label{comparsion}

We compare our model with the baselines and show the comprehensive results in Table \ref{tab:overall}. 
According to the results, deep learning-based methods, such as the GRU-based method (GRU) and CNN-based method (Caser), consistently outperform FPMC in most datasets.
Since such methods leverage more sequential information than FPMC, they can capture more accurate user preference shifts. Self-attention models like Transformer are currently the mainstream in sequential modeling. The self-attention-based sequential recommendation model has more than 5\% performance improvement on most datasets and evaluation metrics than previous methods, which benefits from its ability to model the global context of item-feature correlation. 

Recent temporal sequential modeling methods introduce extra-temporal signals to enhance the temporal correlation among items.
TiSASRec treats different user interaction histories as different time interval sequences, which outperform SASRec on most datasets and evaluation metrics.
SLRS+ introduces the knowledge graph-based relationship and uses the item multi-excitation to enhance the correlation between the target item and user interaction history. Chorus further models the different situations of temporal user-item interaction. KDA proposed the virtual item relation and calculated the relational intensity and frequency embedding for the history items. We can observe that explicit modeling of the relations between items can effectively improve the model's performance. Especially, Chorus and KDA directly contain the temporal item relation loss, which gains significant improvement in all datasets. 

We also conduct experiments on different contrastive learning-based sequential recommendation models. CL4SRec uses item cropping, masking, and reordering as augmentations for contrastive learning, which aims at setting two different views of the same user sequence and maximizing them. S3Rec devises four different pretext tasks to improve the quality of item representations, which performs better than CL4SRec. ContraCL leverages more context information to conduct contrastive learning. DuoRec utilizes both the feature-level dropout masking and the supervised positive sampling to construct contrastive samples. The results actually surpass the traditional SASRec methods especially when $K$ is large. Besides, the category-aware SOTA method DIF-SR outperforms SASRec by at least 5\% on both metrics on most datasets, proving the effectiveness of side information.

Our proposed method combines the dual-transformer module, the dual contrastive learning, and the semantics-enhanced context embedding
module, well estimating the user's hierarchical preference dynamics. Finally, HPM consistently outperforms existing methods on all datasets. The average improvements compared with the best baseline range from 1.53\% to 17.61\% in HR and NDCG. Especially in the Clothing and sports datasets, our model achieves the most impressive improvement, which might be attributed to that they both have higher relational ratios in the test set. The higher relation ratio can provide more stable semantic item relation signals for our SCEL module.

\subsection{Ablation Study}
To verify the effectiveness of different modules in HPM, we comapre the performance of the full model HPM with that of its three variants: 
1) \textit{HPM-S}: HPM without SCEL module.
2) \textit{SPM-O}: We replace the dual transformer structure with a single transformer. Besides, in order to maintain the consistency of the input, we fused the item ID and category information, and then input them into a single transformer.
3) \textit{HPM-C}: HPM without DCL.

\begin{figure}[!h]
\centering
\begin{subfigure}[t]{.5\linewidth}
  \centering	\includegraphics[width=\linewidth,scale=1.00]{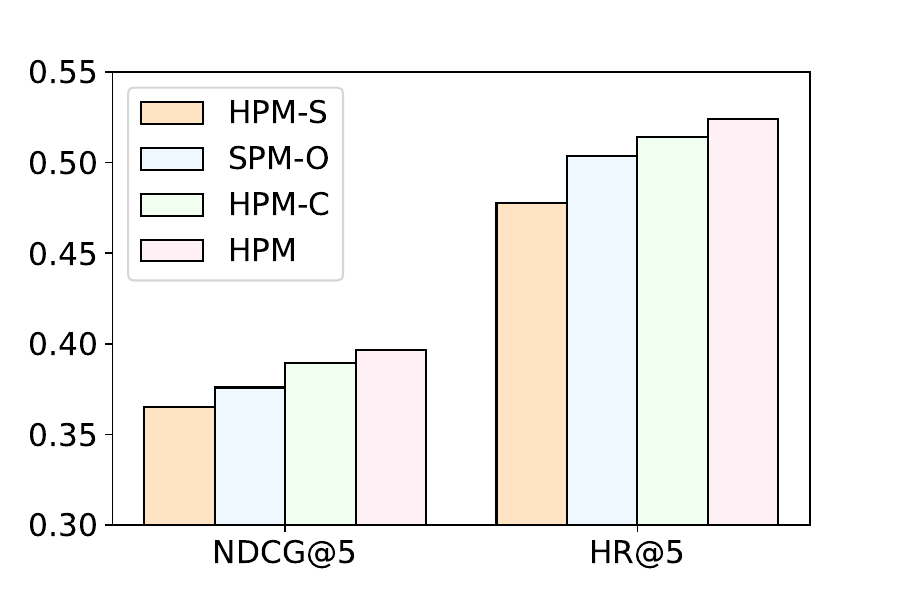}
	\label{fig:ablation1}
\end{subfigure}
\hspace{-5mm} 
\begin{subfigure}[t]{.5\linewidth}
  \centering
 	\includegraphics[width=\linewidth,scale=1.00]{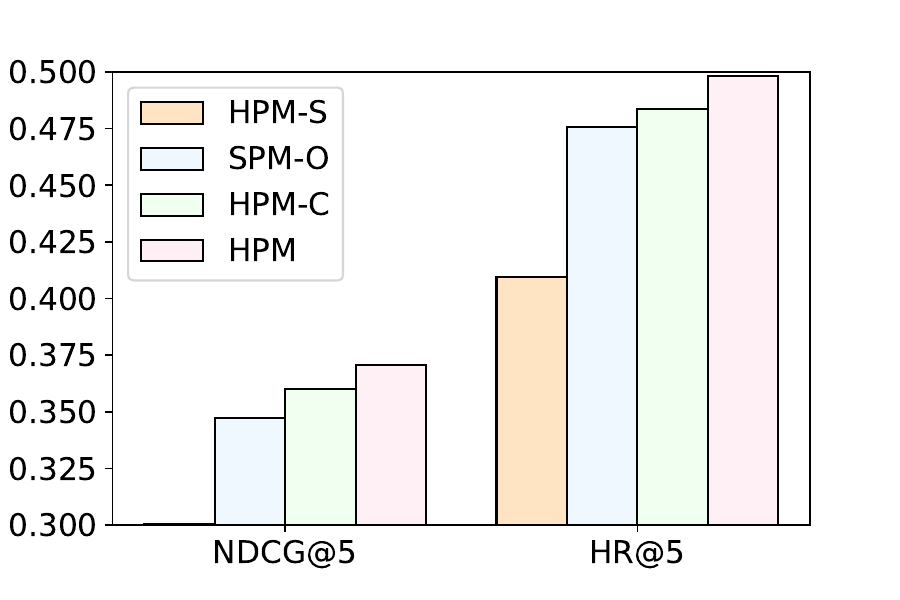}
	\label{fig:ablation2}
\end{subfigure}

\begin{subfigure}[t]{.5\linewidth}
  \centering
 	\includegraphics[width=\linewidth,scale=1.00]{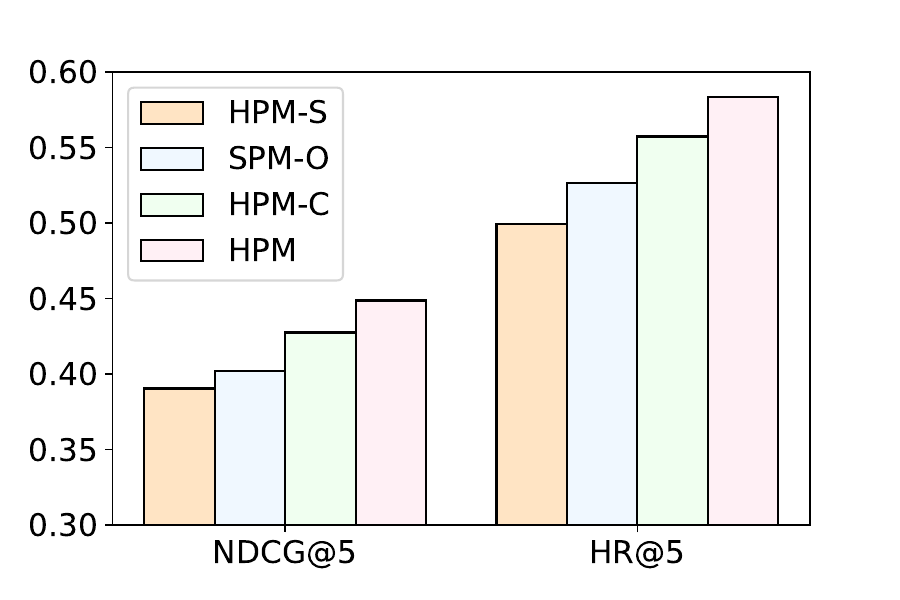}
	\label{fig:ablation3}
\end{subfigure}
\hspace{-5mm} 
\begin{subfigure}[t]{.5\linewidth}
  \centering
 	\includegraphics[width=\linewidth,scale=1.00]{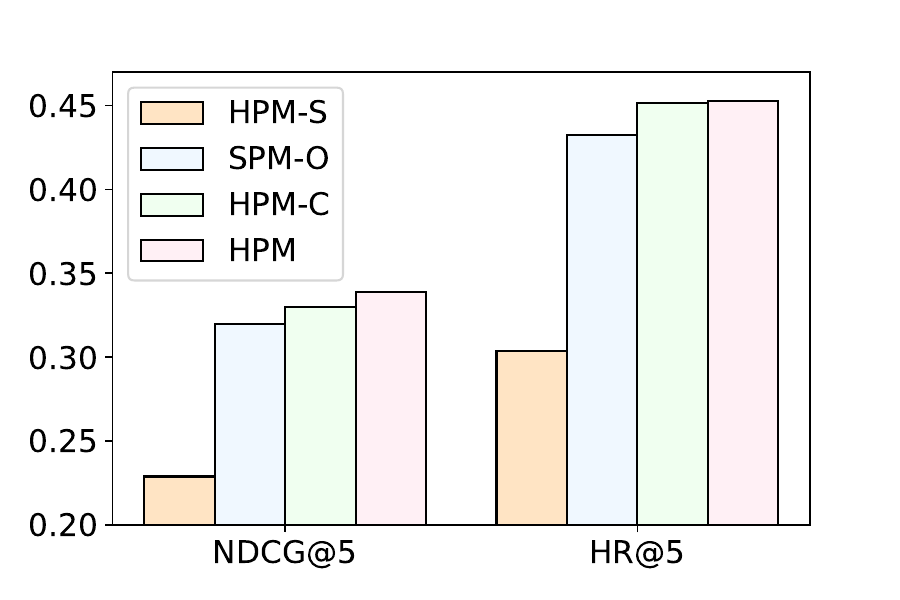}
	\label{fig:ablation4}
\end{subfigure}
\caption{Ablation study of our model (HR@5 and NDCG@5) (Upper left: Beauty, Upper right: Sports, Lower left: Cellphones, Lower right: Clothing).}
\label{fig:ablation}
\end{figure}

As Figure \ref{fig:ablation} illustrated, we conduct the ablation study of our model on four different datasets. First of all, the variant \textit{HPM-S} reports the lowest performance consistently on all datasets, which shows the importance of introducing the explicit relation into our HPM framework. Especially on the sparse Clothing datasets, the SCEL module is particularly important for enhancing the learning of user preferences.
Second, considering our hierarchical preference structure's importance on HPM, we can observe that \textit{SPM-O} results in the loss in all datasets compared to \textit{HPM}. It indicates the significance of the dual transformer to explicitly learn both high- and low-level user preferences. Third, the comparison between \textit{HPM-C} and \textit{HPM} proves the effectiveness of our DCL module. Finally, \textit{HPM} achieves the best results on all four datasets, showing the superiority of our HPM framework.

\subsection{Parameter Sensitivity Test}
\label{PST}

\begin{figure}[!h]
\centering

\begin{subfigure}[t]{.48\linewidth}
  \centering
	\includegraphics[width=\linewidth,scale=1.00]{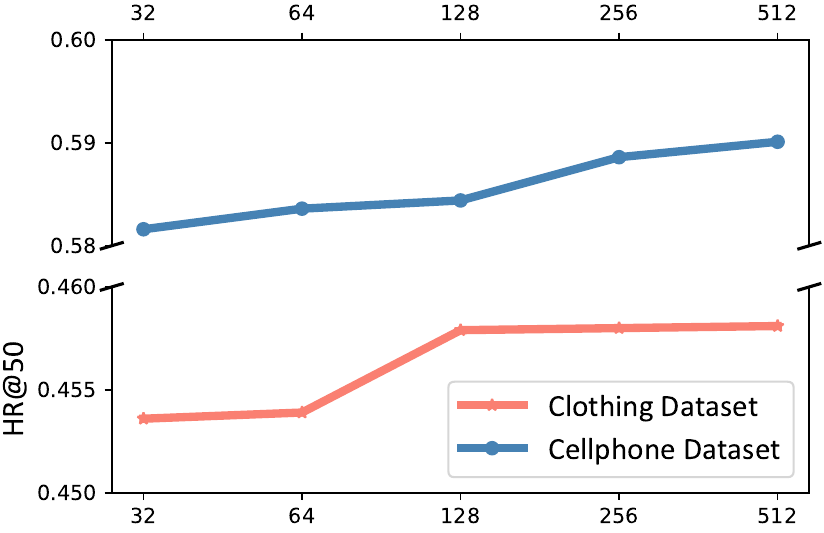}
	\caption{Parameter sensitivity of model embedding size.}
	\label{fig:Embedding_Size}
\end{subfigure}
\hfill
\begin{subfigure}[t]{.48\linewidth}
  \centering
 	\includegraphics[width=\linewidth,scale=1.00]{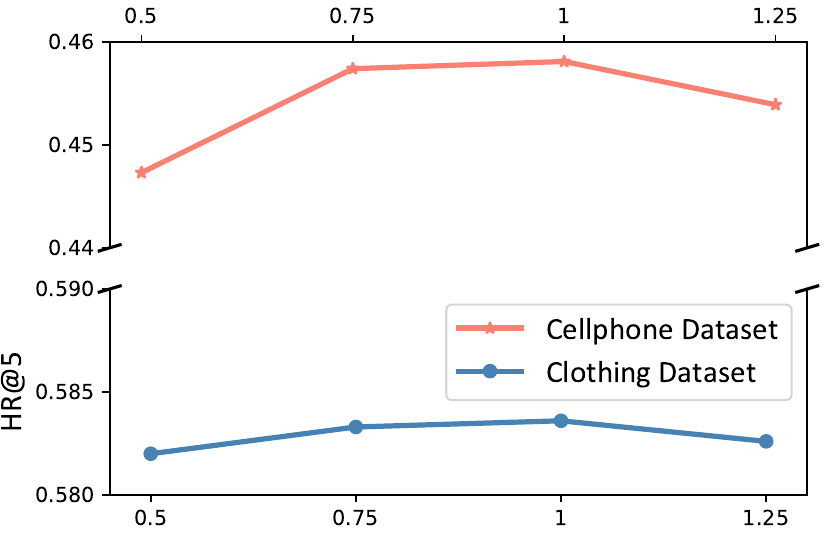}
	\caption{Parameter sensitivity of $\lambda$.}
	\label{fig:coefficient}
\end{subfigure}
\hfill

\caption{Parameter setting's effect on the model performance. (HR@5) on Amazon Clothing dataset.}
\label{fig_comparison3}
\end{figure}

In order to evaluate the effect of different hyper-parameters on the model performance, we conduct parameter sensitivity experiments with ContraCL on the Clothing and Cellphone dataset. We first evaluate the impact of different sizes of the model embedding size, varying from 32 to 512, the $\lambda$=1 and $batch\_size$=64 are fixed. As Figure \ref{fig_comparison3} shows, with the increase of the model embedding size, both of them gain the corresponding improvement. Then we fix the $embedding\_size$=64, $batch\_size$=64 to figure out how contrastive loss coefficient $\lambda$ affects the model performance. The results in Figure \subref{fig:coefficient}) illustrate the model shows the upward trend from 0.5 to 1 then decreases, achieving the best performance when the coefficient $\lambda=1$, indicating the intensity of $\mathcal{L}_{rec}$ and $\mathcal{L}_{cl}$ should be balanced. 
Finally, we check the impact of batch size on the model performance. 
The performance of the model improves with the increase in embedding size.
A similar phenomenon was observed on batch size,
as a larger batch size provides more diverse negative contrastive samples.

\begin{figure}[!h]
\centering

\begin{subfigure}[t]{.48\linewidth}
  \centering
	\includegraphics[width=\linewidth,scale=1.00]{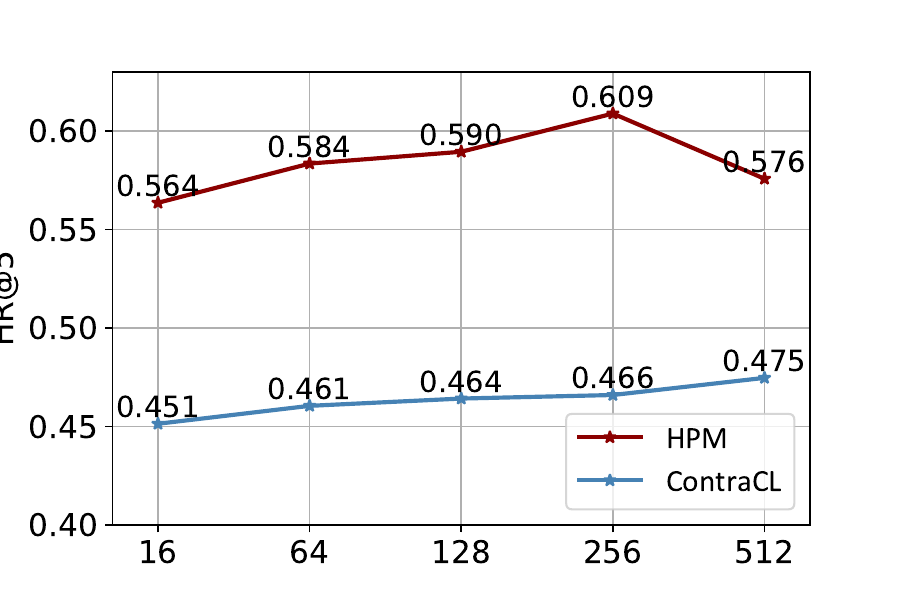}
	\caption{HR@5 comparison w.r.t. Batch Size on Cellphone.}
	\label{fig:batch_size1}
\end{subfigure}
\hfill
\begin{subfigure}[t]{.48\linewidth}
  \centering
 	\includegraphics[width=\linewidth,scale=1.00]{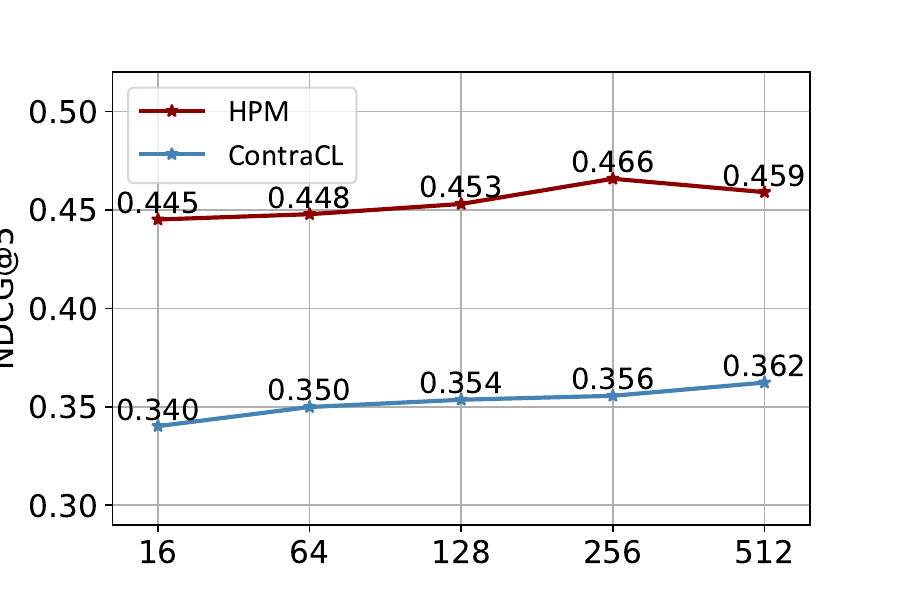}
	\caption{NDCG@5 comparison w.r.t. Batch Size on Cellphone.}
	\label{fig:batch_size2}
\end{subfigure}
\hfill
\caption{Parameter setting's effect on the model performance. (HR@5 and NDCG@5) on Amazon Cellphone dataset.}
\label{fig_comparison2}
\end{figure}

\section{Conclusion}
In this paper, to effectively model the hierarchical preference dynamics of users in the sequential recommendation, which has not been well addressed in the literature, we design a novel hierarchical preference modeling framework called HPM. HPM contains three well-designed modules: the Dual Transformer (DT) module, the Dual-Contrastive Learning (DCL) module, and the Semantics-enhanced Context Embedding Learning (SCEL) module, which work collaboratively to comprehensively learn users' preference dynamics in both item- and category-level. Extensive experiments analyses on real-world datasets verify the superiority of our model over state-of-the-art methods and the rationality of our design. In the future, we will explore more effective methods for discriminatively modeling the different preference-changing patterns at different levels.    

\bibliographystyle{ACM-Reference-Format}
\balance
\bibliography{sample-base}

\end{document}